\documentclass[useAMS,usenatbib]{mn2e}
\usepackage{graphicx}
\usepackage[fleqn]{amsmath}

\def\ie{i.e.}
\def\eg{e.g.}

\title[Gaussian processes for modelling systematics]{A Gaussian process framework for modelling instrumental systematics: application to transmission spectroscopy}
\author[N. P. Gibson et al.]{
N. P. Gibson$^{1}$\thanks{E-mail: Neale.Gibson@astro.ox.ac.uk},
S. Aigrain$^{1}$,
S. Roberts$^{2}$,
T. M. Evans$^{1}$,
M. Osborne$^{2}$ and \newauthor
F. Pont$^{3}$
\\
$^{1}$Department of Physics, University of Oxford, Denys Wilkinson Building, Keble Road, Oxford OX1 3RH, UK\\
$^{2}$Robotics Research Group, Department of Engineering Science, University of Oxford, Oxford OX1 3PJ, UK\\
$^{3}$School of Physics, University of Exeter, Exeter, EX4 4QL, UK\\
}

\begin{document}

\date{Accepted 2011 September 28.  Received 2011 September 27; in original form 2011 August 8}

\pagerange{\pageref{firstpage}--\pageref{lastpage}} \pubyear{2002}

\maketitle

\label{firstpage}

\begin{abstract}

Transmission spectroscopy, which consists of measuring the wavelength-dependent absorption of starlight by a planet's atmosphere during a transit, is a powerful probe of  atmospheric composition. However, the expected signal is typically orders of magnitude smaller than instrumental systematics, and the results are crucially dependent on the treatment of the latter. In this paper, we propose a new method to infer transit parameters in the presence of systematic noise using Gaussian processes, a technique widely used in the machine learning community for Bayesian regression and classification problems. Our method makes use of auxiliary information about the state of the instrument, but does so in a non-parametric manner, without imposing a specific dependence of the systematics on the instrumental parameters, and naturally allows for the correlated nature of the noise. We give an example application of the method to archival NICMOS transmission spectroscopy of the hot Jupiter HD\,189733, which goes some way towards reconciling the controversy surrounding this dataset in the literature. Finally, we provide an appendix giving a general introduction to Gaussian processes for regression, in order to encourage their application to a wider range of problems.

\end{abstract}


\section{Introduction}

Transiting exoplanets offer a unique opportunity to study the structures and atmospheres of planets outside our Solar System. When a planet passes in front of its parent star as we view it from Earth, it blocks a proportion of starlight which we can use to measure the planet's radius relative to its host. Transmission spectroscopy exploits the fact that the measured radius of a planet is wavelength dependent, as the effective size of a planet's occulting disk depends on the height in the atmosphere at which the planet becomes opaque to starlight, which in turn depends on the atomic and molecular species present in the atmosphere \citep[\eg][]{Seager_2000,Brown_2001}. Thus measuring planetary transits (\ie~the planet-to-star radius ratio) as a function of wavelength allows us to probe the makeup of a planet's atmosphere.

Observations of transmission spectra, in particular using the {\it Hubble Space Telescope} (HST) and {\it Spitzer Space Telescope} (SST) have provided some of the most detailed observations of exoplanet atmospheres to date \citep[see e.g.][]{Charbonneau_2002,Vidal-Madjar_2003,Charbonneau_2005,Deming_2005}. However, extracting transmission spectra from these data sets is by no means trivial, and has led to some discussion regarding the interpretation of these signals. Problems arise because the transmission signal is often dwarfed by instrumental systematics in the datasets. These are caused by imperfect observing conditions, such as changes in the pointing, detector temperature, optics and pixel-to-pixel sensitivity, and result in correlated noise in the light curves. This has led to the development of some novel techniques to correct and remove instrumental systematics, usually by modelling them as a deterministic function of auxiliary measurements from the dataset \citep[see \eg][]{Brown_2001b,Gilliland_2003,Pont_2007}. Often however, the choice of systematics model will significantly affect the transmission spectra, and therefore the detection of atmospheric species. Attempting to develop a more general and robust framework to infer transit parameters in the presence of instrumental systematics and correlated noise provides the motivation for this work.

The problem of correlated noise for observations of transiting exoplanets was first raised by \citet{Pont_2006}, who provided methods to estimate its effects on time series analysis, and take it into account when inferring physical properties of exoplanets from such data sets. Indeed, the problems associated with systematics appear in many different areas of exoplanet observations, including transit timing analysis \citep[\eg][]{Gibson_2009} and radial velocity measurements \citep[\eg][]{Pont_2011}. Notably, \citet{Carter_2009} developed a fast and powerful wavelet based method to model and remove time-correlated noise from transit light curves, without the need for additional inputs. Often however, we have auxiliary information from the data, such as measurements of pointing drifts, or changes in the detector and optical conditions, thought to be the cause of the instrumental systematics. This additional information should ideally be used to model systematic trends if available.

Instrumental systematics have proved to be particularly problematic for HST/NICMOS observations of transmission spectra \citep[\eg][]{Pont_2009, Swain_2008, Gibson_2011}. \citet[][hereafter SVT08]{Swain_2008} observed and analysed the transmission spectra of HD 189733b, and interpreted the results as evidence for CH$_4$ and H$_2$O. Following previous HST analyses, the instrumental systematics in the light curves were modelled as a linear function of `optical state parameters'. The optical state parameters are simply auxiliary measurements made directly from the spectra, such as the position, width and angle of the spectral trace, or other parameters reflecting the state of the detector and optics, such as the temperature and satellite orbital phase. \citet[][hereafter GPA11]{Gibson_2011}, reviewed the evidence for molecular species in this dataset using similar methods to model the systematics. This study concluded that the transmission spectrum (and hence the detection of molecules) was too dependent not only on the functional form of the model, but also on the choice of which parameters to include. However, this was a rather unsatisfactory conclusion, as it did not provide an alternative method to determine the transmission spectrum, and we have since searched for more sophisticated techniques which can provide a robust interpretation of these datasets. This paper introduces the use of Gaussian Process (GP) models to address this problem.

GP models are widely used in the machine learning community \citep{Rasmussen_Williams,Bishop} for Bayesian regression and classification problems, and have recently been adopted in other areas of astrophysics \citep[\eg][]{Way_2006,Mahabal_2008,Way_2009}. Rather than impose a deterministic model, GPs define a distribution over function space. This allows the instrumental systematics to be modelled in a {\it non-parametric} way. Using GPs, we can then marginalise out our ignorance of the systematics model, and determine the posterior distribution of the parameters of interest, in this case the planet-to-star radius ratio. To put it another way, the form of the systematics model is inferred from the data itself, and does not have to be set {\it a priori}. As GPs are a Bayesian technique, they also automatically avoid the problem of over-fitting. As we shall see later, through a sensible choice of prior distributions on our inputs variables, we may also determine which of the input variables are relevant to our dataset.

This paper describes a GP model for inferring transit parameters in the presence of instrumental systematics and correlated noise, when additional parameters are measured from the data such as the optical state parameters mentioned above, but it can easily be generalised to time-correlated noise in the absence of such measurements. We provide a description of the GP model in Sect.~\ref{sect:GPs}, and its application to NICMOS transmission spectroscopy of HD 189733 in Sect.~\ref{sect:application_to_HD189}, followed by our conclusions in Sect.~\ref{sect:discussion}. We also provide an appendix giving a brief introduction to GPs with simple regression examples.

\section{Gaussian Processes for Transmission Spectroscopy}
\label{sect:GPs}

A Gaussian Process (GP) is a non-parametric method for regression, used extensively for regression and classification problems in the machine learning community.  A GP is defined as a collection of random variables, any finite number of which have a joint Gaussian distribution. A brief introduction to GPs is provided in the appendix, along with references for further reading.

Here, we use a GP to specify a non-parametric model of the instrumental systematics for a transit observation, along with a deterministic transit model to infer the transit parameters. We begin by defining the transmission spectroscopy data set, and briefly re-stating the `standard' linear model used in previous analyses, before introducing our GP model.

\subsection{Transmission spectroscopy data sets}
\label{sect:transmission_datasets}

Reduced transmission spectroscopy data sets consist of multiple transit light curves as a function of time. Here, we consider each wavelength channel independently, and store the $N$ flux measurements in the vector $\bmath f = (f_1,\dots,f_N)^T$ observed at time $\bmath t = (t_1,\dots,t_N)^T$. Often, additional parameters (or optical state parameters) are measured which describe the behaviour of the instrument as a function of time. These are given by the vector $\bmath x_n = (x_{n,1},\dots,x_{n,K})^T$ at time $n$, where $K$ is the number of additional parameters. These are collected in the $N \times K$ matrix $\mathbfss{X} = (\bmath{x}_1, \ldots, \bmath{x}_N)^T$, and hereafter are called the {\it input parameters}. As each light curve is treated independently, our model can easily be extended to single passband photometric transit observations, using additional measured inputs, or simply using time as the only input in order to specify time-correlated noise.

\subsection{Linear baseline model}
\label{sect:linear_baseline}

Often, the baseline flux, which is how the light curve would behave in the absence of a transit due to instrumental systematics, is modelled as a linear combination of the input parameters:
\[
\bmath f = \mathbfss{X}\bbeta + \bepsilon,
\]
where $\bbeta = (\beta_1,\dots,\beta_K)^T$ is a vector containing the coefficients of each input parameter, and $\bepsilon  = (\epsilon_1,\dots, \epsilon_N)^T$ are the residuals, assumed to be independent Gaussian noise. The best-fit coefficients are then found by linear least squares for the out-of-transit data, given by
\[
\hat\bbeta = (\mathbfss{X}^T \mathbfss{X})^{-1} \mathbfss{X}^T \bmath f.
\]
The baseline flux is  constructed for all data points using the best-fit coefficients, and the transit light curve is divided through by the baseline model to remove the instrumental systematics. The `decorrelated' light curve is then fitted with a transit model to determine the parameters of interest, using varying methods to propagate the uncertainties in the systematics model.

\subsection{Gaussian process model}
\label{sect:GPmodel}

Rather than impose a simple, deterministic function on the instrumental model, we model each wavelength channel as a GP with a transit mean function:
\[
f(t, \bmath{x}) \sim \mathcal{GP} \left (T(t,\bphi) , \mathbf\Sigma(\bmath x, \btheta)  \right).
\]
Here, $T$ is the transit function depending on $t$ and the transit parameters $\bphi$, modelled using the analytic equations of \citet{Mandel_Agol_2002}. $\mathbf\Sigma$ is the covariance matrix, which is a function of $\bmath x$ and parameters $\btheta$, referred to as {\it hyperparameters} of the GP\footnote{Strictly speaking, the transit parameters are also hyperparameters of the GP, but here we refer to them as the transit parameters for simplicity.}. The definition of a GP means that the joint probability distribution for the finite set of observations $\bmath f$ is a multivariate Gaussian distributed about a transit function $T$, with covariance $\mathbf\Sigma$, given by
\begin{equation}
\label{eq:norm}
p(\bmath f| \mathbfss{X},\btheta,\bphi) = \mathcal{N} \left (T(\bmath t,\bphi) , \mathbf\Sigma(\mathbfss{X},\btheta)  \right).
\end{equation}
The instrumental systematics and noise are specified fully by the covariance matrix, which in turn is specified by a covariance function or {\it kernel}. As the instrumental systematics are a function of the input parameters, we adopt the kernel:
\[
\mathbf\Sigma_{nm} = k(\bmath{x}_n, \bmath{x}_m) = \xi \exp\left[-\sum_{i=1}^K\eta_i(x_{n,i} - x_{m,i})^2\right] + \delta_{nm} \sigma^2.
\]
This returns a scalar covariance for each pair of inputs, defining each element of the covariance matrix, $\mathbf\Sigma_{nm}$. $\xi$ is a hyperparameter that specifies the maximum covariance, and $\boldeta = (\eta_1,\dots,\eta_K)^T$ are the inverse scale parameters for each input vector. To incorporate white noise, we add a variance term $\sigma^2$ to the diagonal of the covariance matrix, where $\delta$ is the Kronecker delta function. We hereafter use the notation $\btheta = (\xi,\boldeta^T,\sigma^2)^T
$ to represent the covariance hyperparameter vector.

The interpretation of this kernel is straightforward; data points that are nearby each other in input space are highly correlated, and data points that are far from each other in input space are relatively uncorrelated. This kernel therefore describes a smooth function of the input parameters, with the addition of white noise. The length scales determine how important each vector in input space is to determine the `closeness' of two data points.
It therefore enables {\it Automatic Relevance Determination} \citep[ARD,][also known as shrinkage or ridge regression]{Neal}, where the $\eta_i$ associated with each input parameter becomes small when the parameter is of little relevance to explain the data set. When $\eta_i \to 0$ the parameter no longer influences the covariance matrix, and does not contribute to the model. In other words, the instrumental systematics model assumes a similar value when all {\it relevant} input parameters are close. This kernel is suitable for application to transmission spectroscopy datasets, as squared exponential kernels are suitable for processes with a dominant length scale in each input dimension. In transmission spectroscopy we are interested in instrumental systematics within a narrow range of timescales, commensurate with the duration of the transit, therefore input parameters that vary in these timescales. Frequencies higher than the sampling can be treated as white noise, and frequencies lower can be modelled using a simple baseline function incorporated into the mean function.

There are other kernel functions available that could be used to model such behaviour, in particular the Mat\'ern class. This can be seen as a `rougher' generalisation of the squared exponential kernel, that results in less smooth functions of the inputs. However, this kernel would require an extra hyperparameter for each input parameter, and would likely prohibit marginalisation over all the hyperparameters, and perhaps allow too much freedom in the instrument model. As the squared exponential kernel defines a prior distribution over smooth function of the inputs, we can further justify the use of the this kernel by noting that this incorporates all common functions typically used in instrument models to account for systematics; for example, models that are linear, quadratic, and higher order functions of the inputs, and thus represents a generalisation of previous work. Indeed, this kernel reflects our prior beliefs about the properties of the data. The systematics are dominated by pixel-to-pixel sensitivity variations, which we would expect to smoothly vary with the inputs, particularly when we consider that each wavelength channel is binned over many pixels. Nonetheless, readers should be aware that there are many covariance kernels available with different properties. The choice of kernel is an important consideration in applying GP models, but the use of a non-parametric systematics model allows much greater flexibility than imposing a deterministic parametric form.

We have now defined a distribution over smooth functions of the input parameters to model the instrumental systematics. The functional form that describes the baseline function may now be inferred from the data itself. As the joint probability of the observations defined by the GP is multivariate Gaussian (Eq.~\ref{eq:norm}), the log marginal likelihood function is written as
\begin{equation}
\label{eq:likelihood}
\log \mathcal{L}(\bmath{r}| \mathbfss{X},\btheta,\bphi) = -\frac{1}{2} \bmath{r}^T\, \mathbf{\Sigma}^{-1} \, \bmath{r} -\frac{1}{2}\log | \mathbf{\Sigma}| -\frac{N}{2} \log\left(2\pi\right),
\end{equation}
where $\mathbf{r}$ is a vector containing the residuals from the mean function:
\[
\bmath{r} = \bmath{f} - T(\bmath{t,\bphi}).
\]
This defines the likelihood as a function of the mean function parameters $\bphi$, which determine the underlying transit light curve, and the hyperparameters $\btheta$, which determine the behaviour of the baseline function and the noise.

In practice, it is convenient to place explicit priors on the maximum covariance hyperparameter and the scale-length hyperparameters (or {\it hyperpriors}), to encourage their values towards zero if they are truly irrelevant to explain the data. In our model we use gamma hyperpriors (with shape parameter of unity), of the form:
\[
p(x) = \frac{1}{l} \exp\left(-x/l\right),
\]
for $x \ge 0$, and $l$ is the length scale of the hyperprior. The log posterior distribution of the transit parameters and hyperparameters is proportional to the sum of the log marginal likelihood and log hyperpriors, hence we add the log hyperpriors to Eq.~\ref{eq:likelihood} as follows:
\begin{equation}
\label{eq:posterior}
\log \mathcal{P}(\btheta,\bphi| \bmath f, \mathbfss{X}) =  \log \, \mathcal{L}(\bmath{r}| \mathbfss{X},\btheta,\bphi) - \frac{\xi}{l_\xi} - \sum_{i=1}^{K} \left(\frac{\eta_i}{l_{i}}\right) + {C},
\end{equation}
for $\xi,\theta_i \geq 0$, $l$ are the scale lengths of the hyperpriors, and $C$ represents additional constant terms. Note that we do not explicitly add hyperpriors for the transit parameters or variance hyperparameter (implying uniform, improper hyperpriors). The hyperprior length scales are set to a large value to ensure the hyperpriors are non-informative. The challenge is now to infer the transit parameters and hyperparameters from the posterior probability distribution.

\subsection{Inferring transit parameters}
\label{sect:inference}

Now we have specified the log posterior as a function of the hyperparameters $\btheta$ and the transit parameters $\bphi$, it is straightforward (in theory) to infer the planet-to-star radius ratio $\rho$ from the data. For example, to find the maximum posterior solution, the log posterior given by Eq.~\ref{eq:posterior} is optimised with respect to the transit parameters and hyperparameters. In a fully Bayesian treatment, we should obtain the posterior distribution for each parameter of interest by marginalising over all the other parameters of our model, \ie~all other mean-function parameters and hyperparameters. In practice this can be done using Monte-Carlo Markov-Chains (MCMC) methods, which are widely used in the exoplanet community to explore the joint posterior probability distribution of multivariate models \citep[see \eg][]{Cameron_2007,Holman_2006,Winn_2008}. However, each time the hyperparameters are changed (at each step in a chain), we must recalculate the covariance matrix and its inverse in order to evaluate the log posterior. This scales badly with the length of the input time series, requiring $\mathcal{O}(N^3)$ operations. This often makes a full marginalisation over all hyperparameters intractable for large data sets, and at best slow to compute. It is therefore worth introducing an approximation known as type-II maximum likelihood.

Type-II maximum likelihood involves fixing all the hyperparameters at their maximum likelihood values, and marginalising over the remaining parameters of interest. In this case we will find the maximum posterior solution, and marginalise over the remaining transit parameters, although we still refer to this procedure as type-II maximum likelihood as in the literature. This is a valid approximation when the posterior distribution is sharply peaked at its maximum around the covariance hyperparameters. First, the log posterior function is optimised with respect to the hyperparameters $\btheta$ \emph{and} transit parameters $\bphi$, in order to determine the maximum posterior solution. This procedure still requires the covariance matrix is inverted, but much fewer times than for a full marginalisation over all parameters. The hyperparameters $\btheta$ are then held fixed at their maximum posterior values, and to obtain the (approximate) posterior distributions for each parameter of interest, one then marginalises over the remaining transit parameters, which does not require a matrix inversion at each step. One caveat is that we don't know if the posterior distribution is sharply peaked until we marginalise over all parameters. Consequently it is always preferable to marginalise over all hyperparameters of the GP when possible, and indeed, this is the final approach used in this work. However, for many sets of data using the same kernel function, we might confirm the effectiveness of type-II maximum likelihood for a particular subset, and apply it to the remainder. Nevertheless, in many cases fixing the covariance hyperparameters might be a better approach than imposing a deterministic model of the systematics, but such approximations must be used with caution. In the following section, we apply our GP model to NICMOS transmission spectroscopy of HD 189733, and use both type-II maximum likelihood and a full marginalisation over all transit parameters and hyperparameters to obtain the transmission spectrum. Of course, results from the full marginalisation are preferred, but the comparison to type-II maximum likelihood is a worthwhile exercise.

\section{Application to NICMOS observations of HD 189733}
\label{sect:application_to_HD189}

\subsection{The dataset}
\label{sect:data}

\begin{figure*}
\includegraphics[width=190mm]{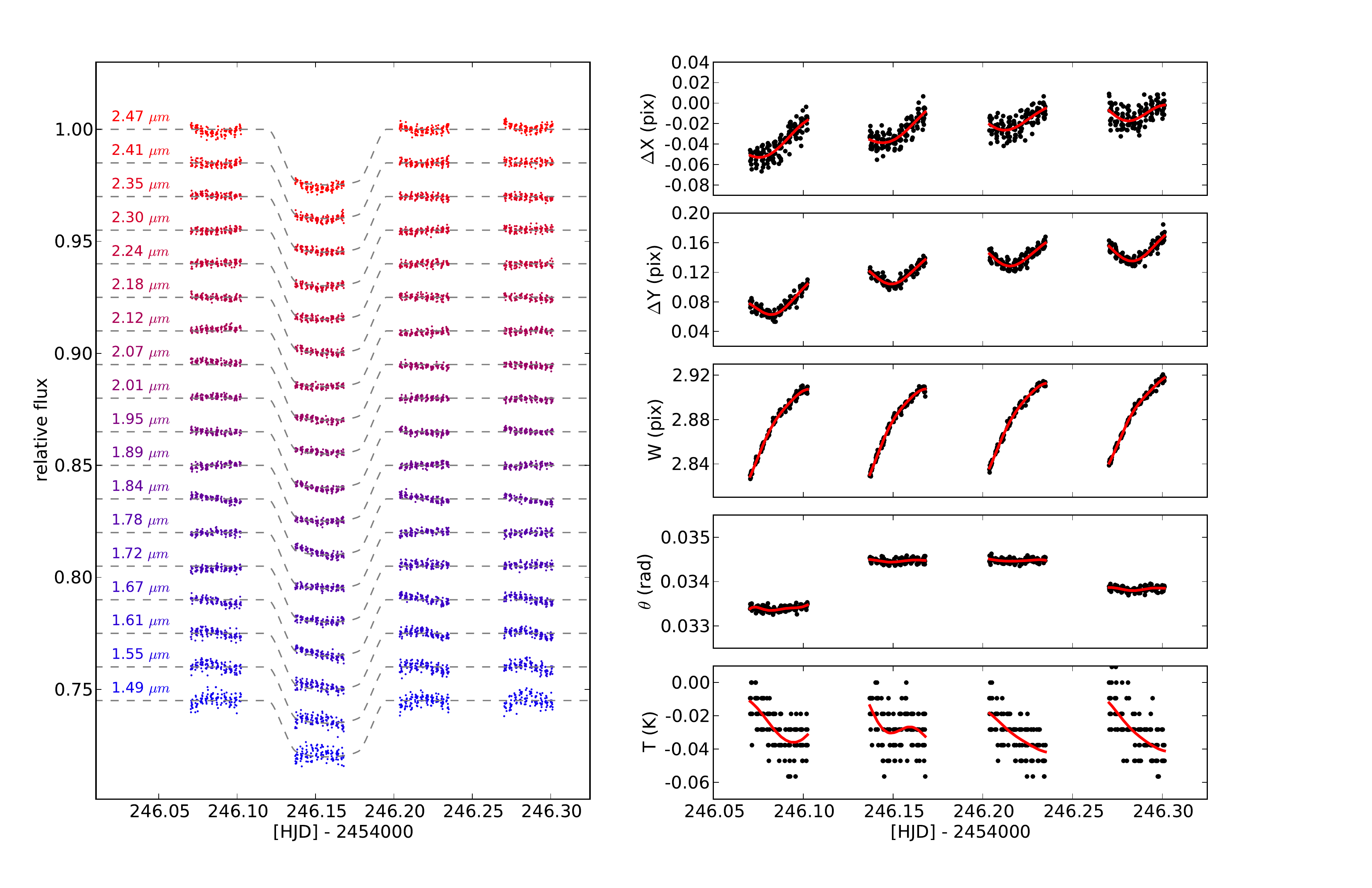}
\caption{The `raw' HD 189733 NICMOS dataset used as an example for our Gaussian process model. Left: Raw light curves of HD 189733 for each of the 18 wavelength channels, from 2.47\micron~to 1.49\micron~top to bottom. Right: The optical state parameters extracted from the spectra plotted as a time series. These are used as the input parameters for our GP model. The red lines represent a GP regression on the input parameters, used to remove the noise and test how this effects the GP model.}
\label{fig:HD189733_raw_data}
\end{figure*}

Here we provide a summary of the HST/NICMOS observations of HD 189733, used as a test case for our GP model. These were originally analysed by SVT08 and reanalysed by GPA11, and the dataset used in this paper is the same as that used in GPA11. For a more detailed account of the observations and data reduction methods, we refer the readers to the aforementioned papers.

A transit of HD 189733 was monitored on 25 May 2007, using the G206 grism covering the wavelength range 1.4 -- 2.5 \micron. As HD 189733 is not in the continuous viewing zone of HST, the light curve was observed over 5 half-orbits, although orbit 1 is excluded as it shows much larger systematic effects attributed to spacecraft `settling'. The light curves for 18 wavelength channels were extracted for 638 spectra, along with optical state parameters. These are the position of the spectral trace along the dispersion axis ($\Delta X$), the average position of the spectral trace along the cross-dispersion axis ($\Delta Y$), the angle the spectral trace makes with the x-axis ($\psi$), and the average width of the spectral trace ($W$). The temperature\footnote{No direct measurement of the temperature of the detector exists, rather a the temperature of the NIC1 mounting cup is used as a proxy measurement, see GPA11 for details.} ($T$), and orbital phase ($\phi_H$) were also determined for each image.

In total, our dataset consists of a time-series of 519 flux measurements (for orbits 2--5 only) in 18 wavelength channels, and time-series of 6 optical state parameters. The 18 light curves and the optical state parameters are shown in Fig.~\ref{fig:HD189733_raw_data}. The light curves show clear instrumental systematics, and these must be accounted for when measuring the transit depth at each wavelength. We have strong reasons to suspect that the systematics in the light curves are related to the optical state parameters in some way. This is a reasonable assumption, as changes in the position, angle and width of the spectrum can obviously have an effect on the flux collected in each wavelength bin, as can the temperature of the detector. However, it is not possible to construct a reliable deterministic physical model which relates these parameters to the baseline flux, as it depends on each individual pixel's sensitivity and response to temperature, as well as other complex changes in the optics.

For each wavelength channel, the GP model outlined in Sect.~\ref{sect:GPs} was applied, where $\bmath f$ contains the 519 flux measurements, and \mathbfss{X} contain the input parameters, where $\bmath x_n = (\Delta X_n,\Delta Y_n,W_n,\psi_n,T_n,\phi_n)^T$. To test that the model was not influenced by the noise of the input parameters, a GP regression model with zero mean function was fitted to each input parameter using a squared exponential kernel, with time as the only input. The hyperparameters were optimised with respect to the likelihood function, and the regression is shown by the red lines in Fig.~\ref{fig:HD189733_raw_data}. The inference described in the following two sections was repeated using these de-noised input parameters\footnote{We did not use the GP smoothed parameters for $T$, as we were not confident the fit reliably represented the underlying structure.}. We found that this had little influence on the results, and report the results using the noisy input parameters. We also checked that the choice of hyperprior length scale had little effect on the results, by setting them to large values, and repeating the procedure with varying length scales ensuring the transmission spectra were not significantly altered.

\subsection{Type-II maximum likelihood}
\label{sect:typeII}

As mentioned in Sect.~\ref{sect:inference}, a useful approximation is Type-II maximum likelihood. First, the log posterior is maximised with respect to the hyperparameters and variable mean-function parameters using a Nelder-Mead simplex algorithm \citep[see \eg][]{NumericalRecipes}. The transit model is the same as that used in GPA11, which uses \citet{Mandel_Agol_2002} models calculated assuming quadratic limb darkening and a circular orbit. All non-variable parameters were fixed to the values given in \citet{Pont_2008}, except for the limb darkening parameters, which were calculated for each wavelength channel \citep[GPA11,][]{Sing_2010}. The only variable mean function variable parameters are the planet-to-star radius ratio, and two parameters that govern a linear baseline model; an out-of-transit flux $f_{oot}$ and (time) gradient $T_{grad}$.

\begin{figure*}
\includegraphics[width=120mm]{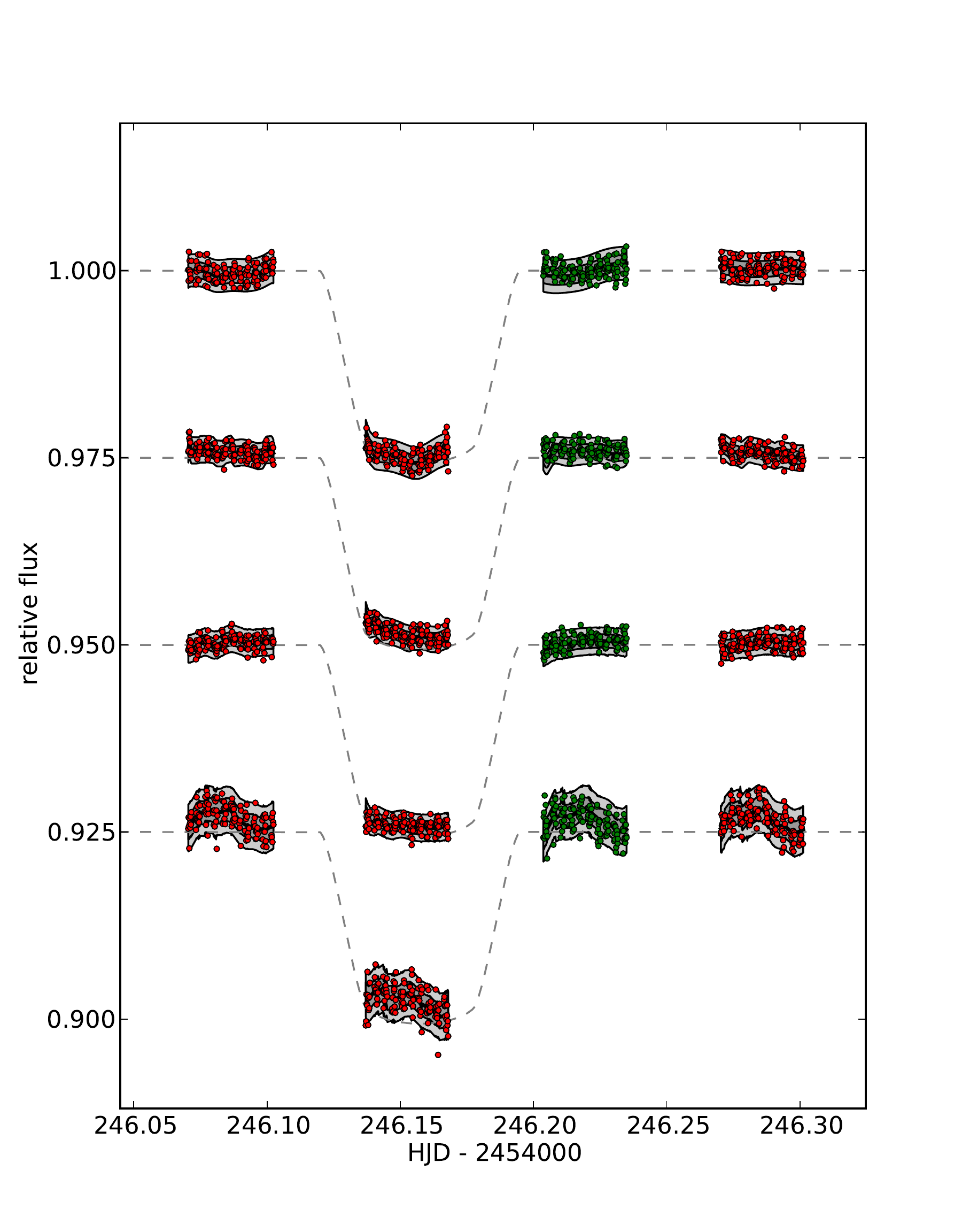}
\caption{Examples of type-II maximum likelihood regression for four light curves of HD 189733. The red points show those data used in the training process (orbits 2, 3 and 5), and the green points show the remaining data (orbit 4). The predictive distributions are shown by the grey regions, which mark the 1 and 2 $\sigma$ confidence intervals. The GP model predictions are consistent with the measured data, indicating that our GP model is an effective model of the instrumental systematics. In practice all four orbits are used to model the light curve and systematics, and infer the planet-to-star radius ratio, hence the systematics model will be even better constrained than shown here.}
\label{fig:TypeII_ML_eg}
\end{figure*}

An example of the predictive distributions found using type-II maximum likelihood is shown in Fig.~\ref{fig:TypeII_ML_eg} for four of the wavelength channels. In this example, only orbits 2, 3 and 5 are used to determine the parameters and hyperparameters of the GP (or `train' the GP), and are shown by the red points\footnote{We used the smoothed hyperparameters here for aesthetic purposes. Using the noisy hyperparameters would simply result in noisier 1 and 2$\sigma$ limits, but with similar structure.}. Orbit 4 (green points) was not used in the training set. Predictive distributions were calculated for orbits 2--5, and are shown by the grey regions, which plot the 1 and 2$\sigma$ confidence intervals. The predictive distribution is a good fit to orbit 4, showing that our GP model is effective at modelling the instrumental systematics. The final systematics model will of course be even more constrained than in this example, as we use orbits 2--5 to simultaneously infer parameters of the GP and transit function, in particular since orbit 4 is the has the most similar input parameters to the in-transit orbit (Fig.~\ref{fig:HD189733_raw_data}).

Now that all parameters and hyperparameters are optimised with respect to the posterior distribution, the hyperparameters are held fixed. This means the inverse covariance matrix and log determinant used to evaluate the log posterior are also fixed, and need to be calculated only once. An MCMC is used to marginalise over the remaining parameters of interest, in this case the planet-to-star radius ratio and the linear baseline coefficients. Our MCMC implementation is adapted from \citet{Gibson_2010}, and uses the log posterior function given by Eq.~\ref{eq:posterior} as the merit function.

For each of the 18 wavelength channels, four separate chains of length 20\,000 were computed, with the starting parameter set initialised to the maximum posterior distribution parameters with small perturbations applied to the variable parameters. During the burn-in phase (first $20\%$ of each chain), the global and relative stepsizes are adapted so that $\sim$25\% of proposal parameters sets are accepted. After burn-in is complete the chains have converged and now sample from the posterior probability distribution. These samples are used to estimate the marginalised posterior distributions for the planet-to-star radius ratio. For each wavelength channel, convergence was confirmed for the four chains by calculating the Gelman \& Rubin (GR) statistic \citep{GelmanRubin_1992}, and in all cases it was within 1\% of unity, a good sign of mixing and convergence.

\begin{figure*}
\includegraphics[width=85mm]{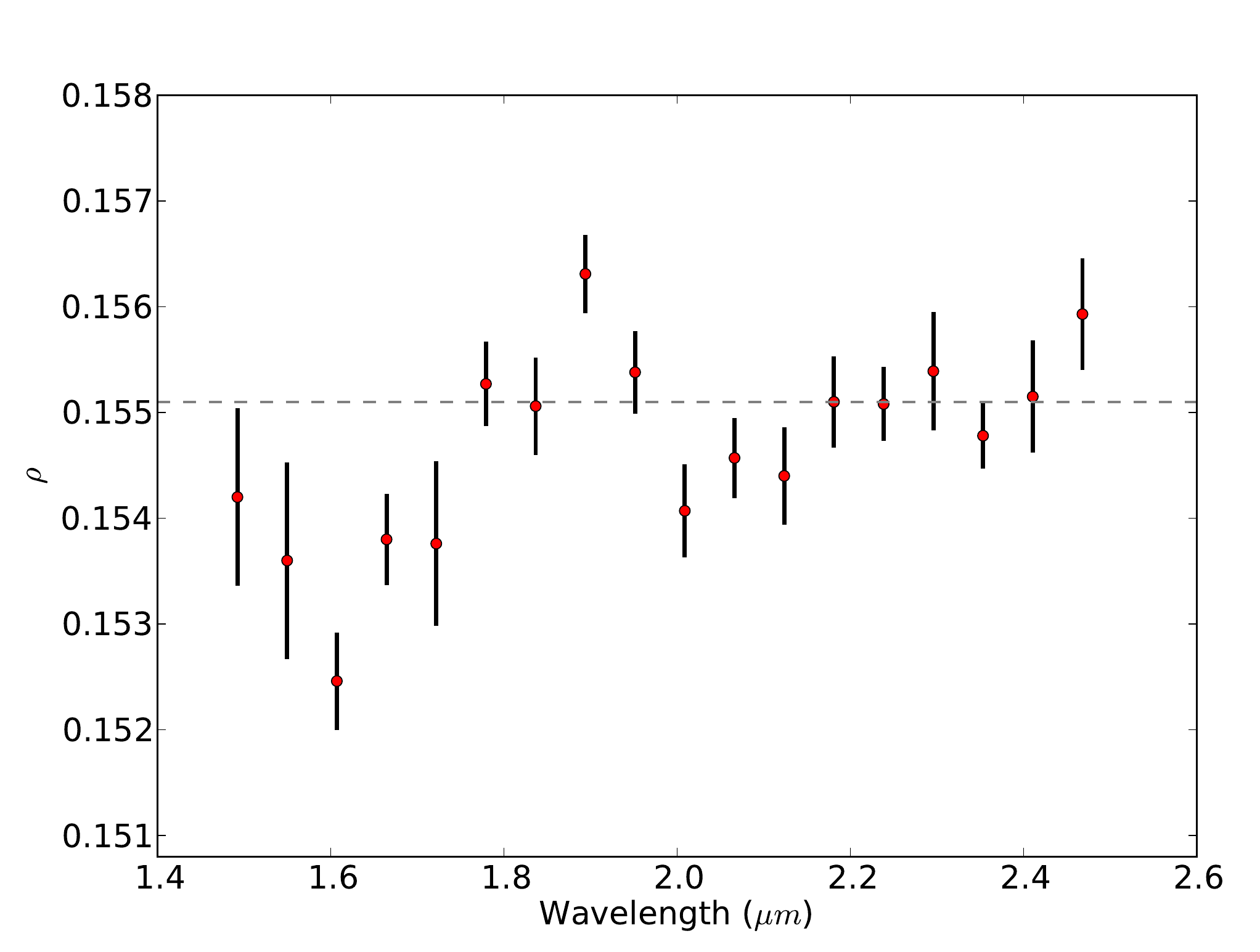}
\includegraphics[width=85mm]{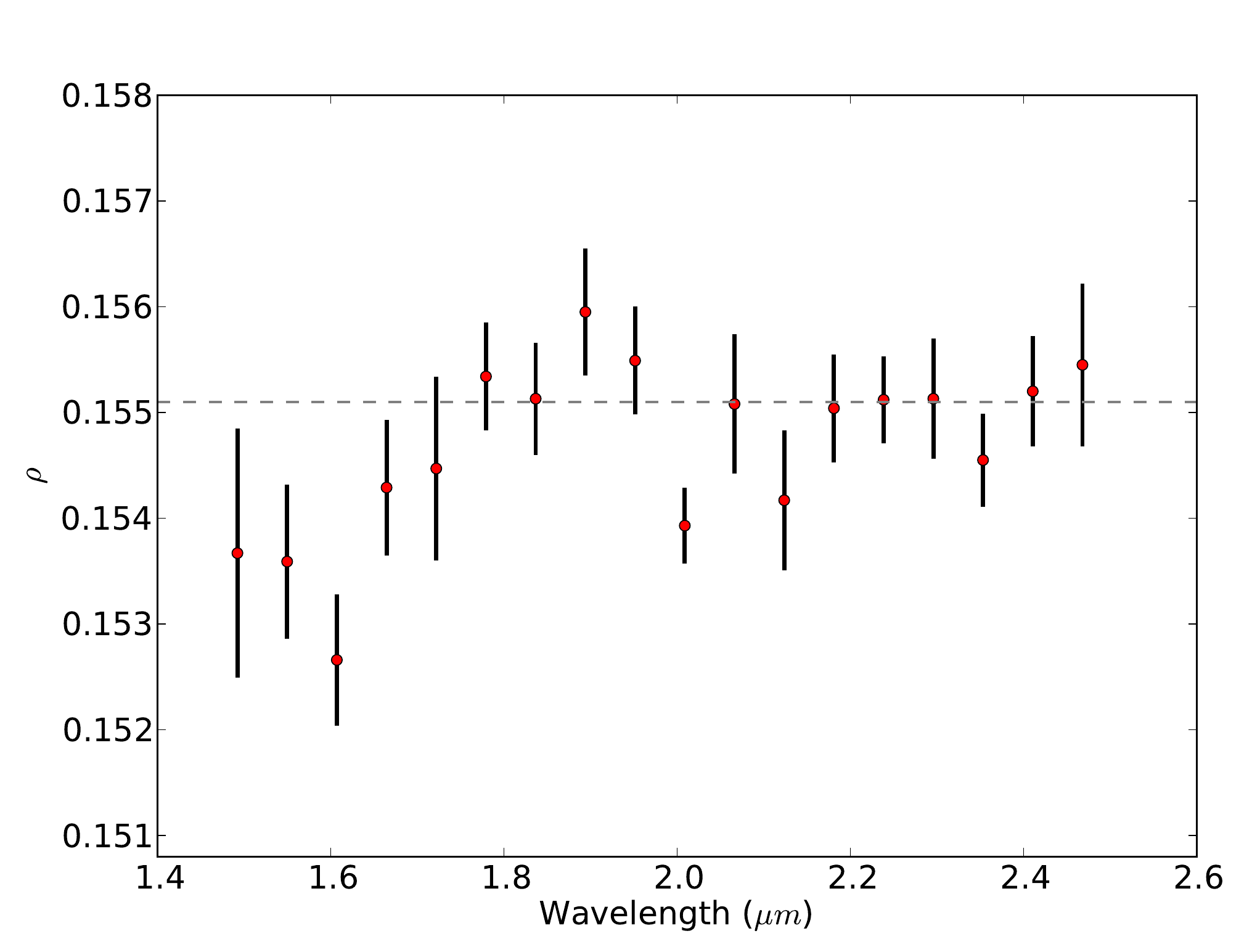}
\caption{NICMOS transmission spectra of HD 189733 produced from the GP model. The horizontal dashed line represents the planet-to-star radius ratio determined from the white light curve. Left: using the maximum likelihood type-II approximation (see Sect.~\ref{sect:typeII}
).  Right: obtained by marginalising over all other parameters and hyperparameters (see Sect.~\ref{sect:marginalisation}).}
\label{fig:spec}
\end{figure*}

The resulting transmission spectrum consists of the planet-to-star radius ratio for each wavelength channel, and is shown in the left plot of Fig.~\ref{fig:spec}. The dashed grey line shows the value of $\rho$ for the `white' light curve (\ie~the light curve given by the sum of all wavelength channels) reported in GPA11.

\subsection{Marginalisation over the hyperparameters}
\label{sect:marginalisation}

Despite the need to invert a matrix at every calculation of the log likelihood function, it is still feasible to marginalise over all the parameters and hyperparameters of the GP model for the HD 189733 dataset using MCMC. The same MCMC routine as described in the previous section is used, this time allowing all the hyperparameters of the GP to vary along with the mean function parameters. The chains were initialised from the maximum posterior values, again with a small perturbation applied to each variable parameter.

\begin{figure*}
\includegraphics[width=184mm]{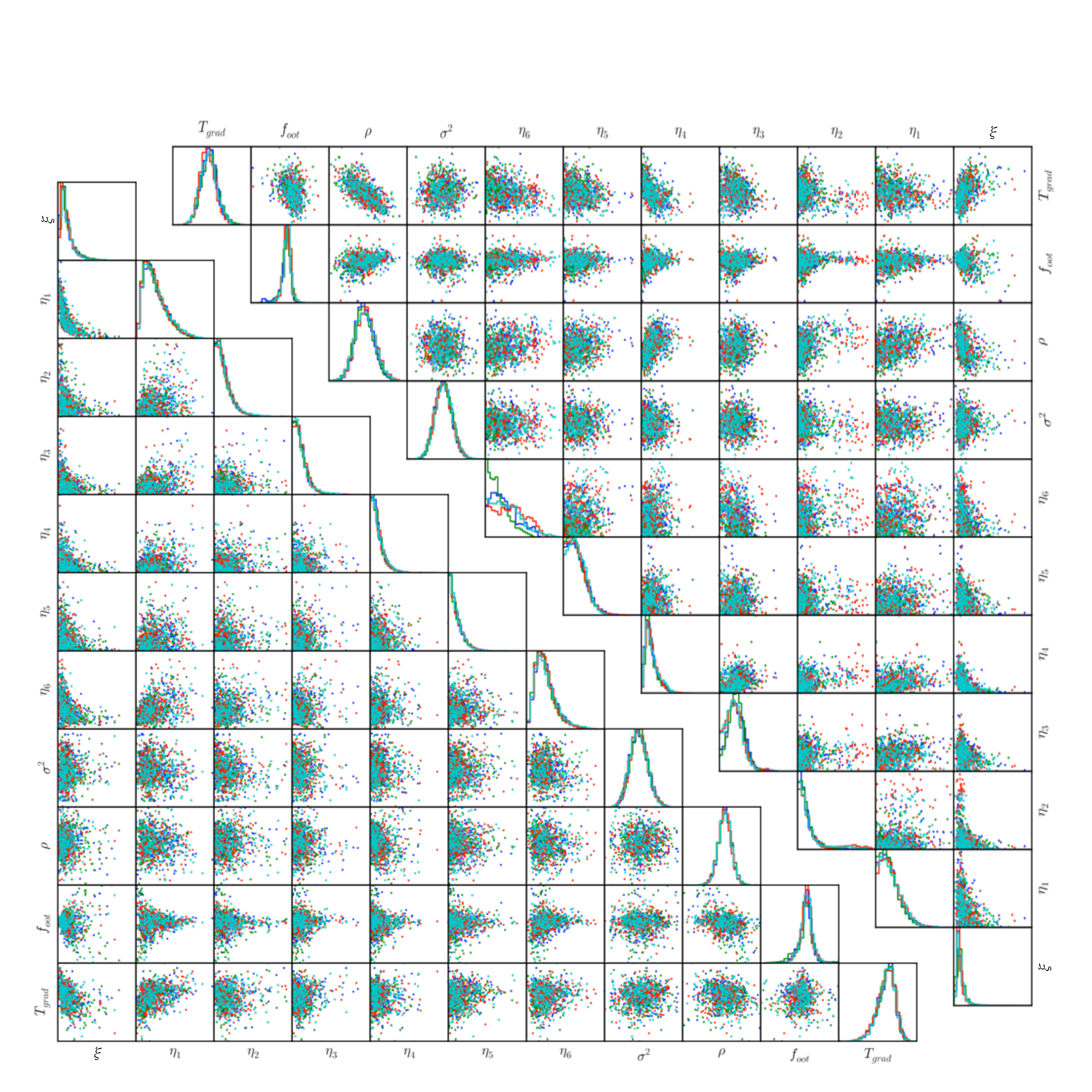}
\caption{Posterior distributions of the variable transit parameters and hyperparameters from the MCMC chains for two of the wavelength channels, shown in the lower and upper triangles. The scatter plots show all pairs of parameters plotted after marginalisation over all other parameters, and the histograms show the marginalised posterior distribution of each individual parameter. The different colours represent the four separate MCMC chains. The lower triangle one shows one of the better converged wavelength channels, and the upper one shows one with some poorly converged hyperparameters.}
\label{fig:MCMC_correlations}
\end{figure*}

For each wavelength channel, four chains of length 150\,000 were calculated. As each evaluation of the posterior probability required a matrix inversion, each chain took about $\sim$2.5 hours to run on a standard desktop computer. In order to run tests in a reasonable time frame, all 72 chains were run in parallel using the Oxford e-Research Centre supercomputing cluster. The chains were again checked for convergence using the GR statistic. For the majority of parameters, convergence within 1\% of unity was achieved. However, some parameters did not converge as efficiently and the GR statistic was as much as $\sim5\%$ from unity. This can be understood in terms of the degeneracy between input parameters. Given the inputs show similar structure, the same overall systematics model may be described using different combinations of scale length parameters. Also, input parameters that have small $\eta$ values are not strongly constrained by the data (or priors) and effectively exhibit random walk behaviour. However, importantly the main parameters of interest $\rho$ did converge for all wavelength channels. Correlation plots for two of the wavelength channels are shown in Fig.~\ref{fig:MCMC_correlations}.

For each of the wavelength channels, different input parameters proved to be relevant to describe the instrument model as inferred from the values of the $\eta$ parameters. The phase is a consistently relevant input parameter for the majority of wavelength channels, and the temperature is not particularly relevant for any channels, likely due to the measurement being particularly noisy and only a proxy measurement. Of course the {\it true} temperature of the detector may still have a significant influence of the light curves, but unfortunately this is not available. The other parameters tend to vary in their influence, in part due to the degeneracy discussed earlier. SVT08 and GPA11 recognised the angle as the most important input parameter. This at first might seem contradictory to our results, but in this case importance was judged on the overall (visual) change in features of the outputted transmission spectrum. Here we judge it on the hyperparameter values from the posterior distribution. Indeed, parameters that are only relevant for a subset of the wavelength channels may in fact have the greatest impact on any features in the transmission spectrum.

The resulting transmission spectrum is shown in the right plot of Fig.~\ref{fig:spec}. Again the dashed line shows the white light curve planet-to-star radius ratio from GPA11. The spectrum is consistent with the type-II maximum likelihood results, the main difference being that the uncertainties are larger. Therefore, in this case it was important to marginalise over all the hyperparameters to obtain good estimates of the uncertainties. Given the convergence was not perfect, we ran the same MCMC procedure several more times. The transmission spectrum and uncertainties produced were almost indistinguishable.

This spectrum is also consistent with the results of SVT08 and GPA11, when using the simple linear basis models. However, the uncertainties are much larger, and do not provide strong evidence for the molecular features reported in SVT08. The central `feature' in the spectrum at $\sim2.1$~\micron~is no longer significant. Bluewards of $\sim1.7$~\micron~the same dip appears, but with lower significance. We emphasise that if the simple linear basis functions provided a good explanation for the systematics, our GP model would have reproduced the same transmission spectrum (and uncertainties), as the GP marginal likelihood function favours long length scales (\ie~slowly varying functions) if they are sufficient to explain the instrumental systematics. Therefore, we strongly encourage anyone wishing to carry out comparisons with theoretical models to use the more conservative, but robust transmission spectrum determined in the present work using the GP model. Not doing so would entail the risk of over-interpreting features in the spectrum. These results are provided in Tab.~\ref{tab:results}.

\begin{table}
\caption{NICMOS transmission spectrum of HD 189733 from our GP model, giving the wavelength, planet-to-star radius ratio $\rho$ and its uncertainty $\Delta\rho$. The wavelengths are displayed red to blue, so that they correspond to the plots in Fig.~\ref{fig:HD189733_raw_data}.}
\label{tab:results}
\begin{tabular}{lcc}
\hline
Wavelength & $\rho$ & $\Delta\rho$\\
($\micron$) & ~ & ~\\
\hline
2.468 & 0.15545 & 0.00077 \\
2.411 & 0.15520 & 0.00052 \\
2.353 & 0.15455 & 0.00044 \\
2.296 & 0.15513 & 0.00057 \\
2.238 & 0.15512 & 0.00041 \\
2.181 & 0.15504 & 0.00051 \\
2.124 & 0.15417 & 0.00066 \\
2.066 & 0.15508 & 0.00066 \\
2.009 & 0.15393 & 0.00036 \\
1.951 & 0.15549 & 0.00051 \\
1.894 & 0.15595 & 0.00060 \\
1.837 & 0.15513 & 0.00053 \\
1.779 & 0.15534 & 0.00051 \\
1.722 & 0.15447 & 0.00087 \\
1.665 & 0.15429 & 0.00064 \\
1.607 & 0.15266 & 0.00062 \\
1.550 & 0.15359 & 0.00073 \\
1.492 & 0.15367 & 0.00118 \\
\hline
\end{tabular}
\end{table}

The dip at the left hand side of the spectrum could indeed be the result of a drop in molecular absorption in the atmosphere. However, we warn the reader that this region of the spectrum is where the flux is lowest, and therefore is susceptible to systematics from poor background subtraction as discussed in GPA11. This has the effect of stretching or compressing the transit light curve in the flux axis, and therefore the dip could be the result of systematics remaining from the data reduction. This raises an important point; GP models, or indeed any sophisticated models, can still suffer from systematic effects not removed from the data sets at the initial reduction stage. A more detailed interpretation of the HD 189733 spectrum will form the subject of a later paper (Gibson et al., in prep), combined with WFC3 near-infrared observations that overlap the blue region of the NICMOS spectrum, and will shed some light on the interpretation of this feature.

\section{Summary and Discussion}
\label{sect:discussion}

We have introduced Gaussian Processes as an alternative to deterministic models for the modelling and removal of systematics in the presence of auxiliary input parameters. GPs provide a powerful Bayesian approach to place distributions over functions. Rather than impose a parametric model of instrumental systematics, by using GPs one can marginalise out their ignorance of the functional form of the systematics model. GPs allow an arbitrary number of additional input parameters, and a framework to determine which parameters are important to the analysis.

We demonstrated the application of our GP model to NICMOS transmission spectroscopy of HD 189733. The transmission spectrum is consistent with those found in SVT08 and GPA11 using simple linear basis functions, although with larger uncertainties. This is because the linear basis functions are not sufficient to explain the instrumental systematics, and therefore do not provide realistic treatment of the uncertainties. Detailed interpretation of these results will be included in a later paper.

Our GP model may also be used in the absence of auxiliary information, by using a time dependant covariance kernel to model time-correlated noise in exoplanet transits. However, due to the considerable computation time required, existing methods, in particular the wavelet method of \citet{Carter_2009}, are considerably faster where applicable for large transit data sets, and are preferable in the absence of additional inputs, although limited to regularly sampled data. Comparing results of both methods for time-correlated noise will provide useful tests, and may form the subject of future work.

Recently, \citet{Waldmann_2011} proposed another non-parametric method for removing systematics from transmission spectroscopy datasets, based on Independent Component Analysis. This uses a completely different approach, and searches for common signals in multiple transit light curves, used to remove the systematics from the data. These new methods based on non-parametric models should provide effective new tools to robustly extract signals in the presence of unknown systematic noise, and the development of such methods are important to future exoplanet research. It would be interesting to compare results from our GP model to those using the approach proposed by \citet{Waldmann_2011}, as the two methods may prove complementary, given that they make different assumptions about the systematics.

GP models provide a general framework that can be applied to many different problems in regression, interpolation and prediction where a deterministic function is not available. The major limitation to applying GP models is that each evaluation of the posterior probability requires a matrix inversion, which scales badly with the size of the data set. This is unavoidable for datasets which require marginalisation over function space. This difficulty may be somewhat eased in the near future by using sparse GP methods \cite[e.g.][]{quinonero2005unifying,walder2008sparse},
which approximate the covariance matrix as sparse, or using utilising rank-reduction methods for large matrices where appropriate \citep[e.g.][]{Foster_2009,Way_2009}, and/or by using GPUÕs to vastly speed up matrix inversion \citep[e.g.][]{Volkov_2008}.
Indeed, ongoing research in this area should open up the application of GPs to many more astronomical datasets.

\section*{Acknowledgments}

All of the data presented in this paper were obtained from the Multimission Archive at the Space Telescope Science Institute (MAST). STScI is operated by the Association of Universities for Research in Astronomy, Inc., under NASA contract NAS5-26555. Support for MAST for non-HST data is provided by the NASA Office of Space Science via grant NNX09AF08G and by other grants and contracts. N. P. G and S. A. acknowledge support from STFC grant ST/G002266/2. The authors would like to acknowledge the use of the Oxford Supercomputing Centre (OSC) in carrying out this work.

\bibliography{../MyBibliography} 

\begin{thebibliography}{37}
\expandafter\ifx\csname natexlab\endcsname\relax\def\natexlab#1{#1}\fi

\bibitem[{{Bishop}(2006)}]{Bishop}
{Bishop} C.~M., 2006, {Pattern Recognition and Machine Learning}. {Springer}

\bibitem[{{Brown}(2001)}]{Brown_2001}
{Brown} T.~M., 2001, \apj, 553, 1006

\bibitem[{{Brown} {et~al.}(2001){Brown}, {Charbonneau}, {Gilliland}, {Noyes},
  \& {Burrows}}]{Brown_2001b}
{Brown} T.~M., {Charbonneau} D., {Gilliland} R.~L., {Noyes} R.~W., {Burrows}
  A., 2001, \apj, 552, 699

\bibitem[{{Carter} \& {Winn}(2009)}]{Carter_2009}
{Carter} J.~A., {Winn} J.~N., 2009, \apj, 704, 51

\bibitem[{{Charbonneau} {et~al.}(2005){Charbonneau}, {Allen}, {Megeath},
  {Torres}, {Alonso}, {Brown}, {Gilliland}, {Latham}, {Mandushev}, {O'Donovan},
  \& {Sozzetti}}]{Charbonneau_2005}
{Charbonneau} D., {Allen} L.~E., {Megeath} S.~T., {Torres} G., {Alonso} R.,
  {Brown} T.~M., {Gilliland} R.~L., {Latham} D.~W., {Mandushev} G., {O'Donovan}
  F.~T., {Sozzetti} A., 2005, \apj, 626, 523

\bibitem[{{Charbonneau} {et~al.}(2002){Charbonneau}, {Brown}, {Noyes}, \&
  {Gilliland}}]{Charbonneau_2002}
{Charbonneau} D., {Brown} T.~M., {Noyes} R.~W., {Gilliland} R.~L., 2002, \apj,
  568, 377

\bibitem[{{Collier Cameron} {et~al.}(2007){Collier Cameron}, {Wilson}, {West},
  {Hebb}, {Wang}, {Aigrain}, {Bouchy}, {Christian}, {Clarkson}, {Enoch},
  {Esposito}, {Guenther}, {Haswell}, {H{\'e}brard}, {Hellier}, {Horne},
  {Irwin}, {Kane}, {Loeillet}, {Lister}, {Maxted}, {Mayor}, {Moutou}, {Parley},
  {Pollacco}, {Pont}, {Queloz}, {Ryans}, {Skillen}, {Street}, {Udry}, \&
  {Wheatley}}]{Cameron_2007}
{Collier Cameron} A., {Wilson} D.~M., {West} R.~G., {Hebb} L., {Wang} X.-B.,
  {Aigrain} S., {Bouchy} F., {Christian} D.~J., {Clarkson} W.~I., {Enoch} B.,
  {Esposito} M., {Guenther} E., {Haswell} C.~A., {H{\'e}brard} G., {Hellier}
  C., {Horne} K., {Irwin} J., {Kane} S.~R., {Loeillet} B., {Lister} T.~A.,
  {Maxted} P., {Mayor} M., {Moutou} C., {Parley} N., {Pollacco} D., {Pont} F.,
  {Queloz} D., {Ryans} R., {Skillen} I., {Street} R.~A., {Udry} S., {Wheatley}
  P.~J., 2007, \mnras, 380, 1230

\bibitem[{{Deming} {et~al.}(2005){Deming}, {Seager}, {Richardson}, \&
  {Harrington}}]{Deming_2005}
{Deming} D., {Seager} S., {Richardson} L.~J., {Harrington} J., 2005, \nat, 434,
  740

\bibitem[{Foster {et~al.}(2009)Foster, Waagen, Aijaz, Hurley, Luis, Rinsky,
  Satyavolu, Way, Gazis, \& Srivastava}]{Foster_2009}
Foster L., Waagen A., Aijaz N., Hurley M., Luis A., Rinsky J., Satyavolu C.,
  Way M.~J., Gazis P., Srivastava A., 2009, J. Mach. Learn. Res., 10, 857

\bibitem[{Garnett {et~al.}(2010)Garnett, Osborne, Reece, Rogers, \&
  Roberts}]{Garnett_2010}
Garnett R., Osborne M., Reece S., Rogers A., Roberts S., 2010, The Computer
  Journal, 53, 1430

\bibitem[{{Gelman} \& {Rubin}(1992)}]{GelmanRubin_1992}
{Gelman} A., {Rubin} D.~B., 1992, Stat. Sci., 7, 457

\bibitem[{{Gibson} {et~al.}(2009){Gibson}, {Pollacco}, {Simpson}, {Barros},
  {Joshi}, {Todd}, {Keenan}, {Skillen}, {Benn}, {Christian}, {Hrudkov{\'a}}, \&
  {Steele}}]{Gibson_2009}
{Gibson} N.~P., {Pollacco} D., {Simpson} E.~K., {Barros} S., {Joshi} Y.~C.,
  {Todd} I., {Keenan} F.~P., {Skillen} I., {Benn} C., {Christian} D.,
  {Hrudkov{\'a}} M., {Steele} I.~A., 2009, \apj, 700, 1078

\bibitem[{{Gibson} {et~al.}(2010){Gibson}, {Pollacco}, {Barros}, {Benn},
  {Christian}, {Hrudkov{\'a}}, {Joshi}, {Keenan}, {Simpson}, {Skillen},
  {Steele}, \& {Todd}}]{Gibson_2010}
{Gibson} N.~P., {Pollacco} D.~L., {Barros} S., {Benn} C., {Christian} D.,
  {Hrudkov{\'a}} M., {Joshi} Y.~C., {Keenan} F.~P., {Simpson} E.~K., {Skillen}
  I., {Steele} I.~A., {Todd} I., 2010, \mnras, 401, 1917

\bibitem[{{Gibson} {et~al.}(2011){Gibson}, {Pont}, \& {Aigrain}}]{Gibson_2011}
{Gibson} N.~P., {Pont} F., {Aigrain} S., 2011, \mnras, 411, 2199

\bibitem[{{Gilliland} \& {Arribas}(2003)}]{Gilliland_2003}
{Gilliland} R.~L., {Arribas} S., 2003, Instrument Science Report NICMOS
  2003-001

\bibitem[{{Holman} {et~al.}(2006){Holman}, {Winn}, {Latham}, {O'Donovan},
  {Charbonneau}, {Bakos}, {Esquerdo}, {Hergenrother}, {Everett}, \&
  {P{\'a}l}}]{Holman_2006}
{Holman} M.~J., {Winn} J.~N., {Latham} D.~W., {O'Donovan} F.~T., {Charbonneau}
  D., {Bakos} G.~A., {Esquerdo} G.~A., {Hergenrother} C., {Everett} M.~E.,
  {P{\'a}l} A., 2006, \apj, 652, 1715

\bibitem[{{Mahabal} {et~al.}(2008){Mahabal}, {Djorgovski}, {Williams}, {Drake},
  {Donalek}, {Graham}, {Moghaddam}, {Turmon}, {Jewell}, {Khosla}, \&
  {Hensley}}]{Mahabal_2008}
{Mahabal} A., {Djorgovski} S.~G., {Williams} R., {Drake} A., {Donalek} C.,
  {Graham} M., {Moghaddam} B., {Turmon} M., {Jewell} J., {Khosla} A., {Hensley}
  B., 2008, in American Institute of Physics Conference Series, Vol. 1082,
  American Institute of Physics Conference Series, {C.~A.~L.~Bailer-Jones},
  ed., pp. 287--293

\bibitem[{{Mandel} \& {Agol}(2002)}]{Mandel_Agol_2002}
{Mandel} K., {Agol} E., 2002, \apjl, 580, L171

\bibitem[{{Neal}(1996)}]{Neal}
{Neal} R.~M., 1996, {Bayesian Learning for Neural Networks}. {Springer}

\bibitem[{{Pont} {et~al.}(2011){Pont}, {Aigrain}, \& {Zucker}}]{Pont_2011}
{Pont} F., {Aigrain} S., {Zucker} S., 2011, \mnras, 411, 1953

\bibitem[{{Pont} {et~al.}(2009){Pont}, {Gilliland}, {Knutson}, {Holman}, \&
  {Charbonneau}}]{Pont_2009}
{Pont} F., {Gilliland} R.~L., {Knutson} H., {Holman} M., {Charbonneau} D.,
  2009, \mnras, 393, L6

\bibitem[{{Pont} {et~al.}(2007){Pont}, {Gilliland}, {Moutou}, {Charbonneau},
  {Bouchy}, {Brown}, {Mayor}, {Queloz}, {Santos}, \& {Udry}}]{Pont_2007}
{Pont} F., {Gilliland} R.~L., {Moutou} C., {Charbonneau} D., {Bouchy} F.,
  {Brown} T.~M., {Mayor} M., {Queloz} D., {Santos} N., {Udry} S., 2007, \aap,
  476, 1347

\bibitem[{{Pont} {et~al.}(2008){Pont}, {Knutson}, {Gilliland}, {Moutou}, \&
  {Charbonneau}}]{Pont_2008}
{Pont} F., {Knutson} H., {Gilliland} R.~L., {Moutou} C., {Charbonneau} D.,
  2008, \mnras, 385, 109

\bibitem[{{Pont} {et~al.}(2006){Pont}, {Zucker}, \& {Queloz}}]{Pont_2006}
{Pont} F., {Zucker} S., {Queloz} D., 2006, \mnras, 373, 231

\bibitem[{{Press} {et~al.}(1992){Press}, {Teukolsky}, {Vetterling}, \&
  {Flannery}}]{NumericalRecipes}
{Press} W.~H., {Teukolsky} S.~A., {Vetterling} W.~T., {Flannery} B.~P., 1992,
  {Numerical recipes in C. The art of scientific computing}, {Press, W.~H.,
  Teukolsky, S.~A., Vetterling, W.~T., \& Flannery, B.~P. }, ed.

\bibitem[{Qui{\~n}onero-Candela \& Rasmussen(2005)}]{quinonero2005unifying}
Qui{\~n}onero-Candela J., Rasmussen C., 2005, The Journal of Machine Learning
  Research, 6, 1939

\bibitem[{{Rasmussen} \& {Williams}(2006)}]{Rasmussen_Williams}
{Rasmussen} C.~E., {Williams} K.~I., 2006, {Gaussian Processes for Machine
  Learning}. {The MIT Press}

\bibitem[{{Seager} \& {Sasselov}(2000)}]{Seager_2000}
{Seager} S., {Sasselov} D.~D., 2000, \apj, 537, 916

\bibitem[{{Sing}(2010)}]{Sing_2010}
{Sing} D.~K., 2010, \aap, 510, A21+

\bibitem[{{Swain} {et~al.}(2008){Swain}, {Vasisht}, \& {Tinetti}}]{Swain_2008}
{Swain} M.~R., {Vasisht} G., {Tinetti} G., 2008, \nat, 452, 329

\bibitem[{{Vidal-Madjar} {et~al.}(2003){Vidal-Madjar}, {Lecavelier des Etangs},
  {D{\'e}sert}, {Ballester}, {Ferlet}, {H{\'e}brard}, \&
  {Mayor}}]{Vidal-Madjar_2003}
{Vidal-Madjar} A., {Lecavelier des Etangs} A., {D{\'e}sert} J.-M., {Ballester}
  G.~E., {Ferlet} R., {H{\'e}brard} G., {Mayor} M., 2003, \nat, 422, 143

\bibitem[{Volkov \& Demmel(2008)}]{Volkov_2008}
Volkov V., Demmel J., 2008, Lu, qr and cholesky factorizations using vector
  capabilities of gpus. Tech. Rep. UCB/EECS-2008-49, EECS Department,
  University of California, Berkeley

\bibitem[{Walder {et~al.}(2008)Walder, Kim, \&
  Sch{\"o}lkopf}]{walder2008sparse}
Walder C., Kim K., Sch{\"o}lkopf B., 2008, in Proceedings of the 25th
  international conference on Machine learning, ACM, pp. 1112--1119

\bibitem[{{Waldmann}(2011)}]{Waldmann_2011}
{Waldmann} I.~P., 2011, ArXiv e-prints

\bibitem[{{Way} {et~al.}(2009){Way}, {Foster}, {Gazis}, \&
  {Srivastava}}]{Way_2009}
{Way} M.~J., {Foster} L.~V., {Gazis} P.~R., {Srivastava} A.~N., 2009, \apj,
  706, 623

\bibitem[{{Way} \& {Srivastava}(2006)}]{Way_2006}
{Way} M.~J., {Srivastava} A.~N., 2006, \apj, 647, 102

\bibitem[{{Winn} {et~al.}(2008){Winn}, {Holman}, {Torres}, {McCullough},
  {Johns-Krull}, {Latham}, {Shporer}, {Mazeh}, {Garcia-Melendo}, {Foote},
  {Esquerdo}, \& {Everett}}]{Winn_2008}
{Winn} J.~N., {Holman} M.~J., {Torres} G., {McCullough} P., {Johns-Krull} C.,
  {Latham} D.~W., {Shporer} A., {Mazeh} T., {Garcia-Melendo} E., {Foote} C.,
  {Esquerdo} G., {Everett} M., 2008, \apj, 683, 1076

\end{thebibliography}
\bibliographystyle{mn2e_astronat} 

\appendix
\section{Gaussian processes for regression}
\label{app:GaussianProcesses}

Gaussian process (GP) models are extensively used in the machine learning community for Bayesian inference in non-parametric regression and classification problems. They can be seen as an extension of kernel regression to probabilistic models. This appendix aims to give a brief introduction to GPs used for regression problems. Our explanations are based on the textbooks by \citet{Rasmussen_Williams} and \citet{Bishop}, where the interested reader will find more complete and detailed information.

\subsection{Introducing Gaussian Processes}

We first define a collection of $N$ observations, consisting of independent variables $\bmath{x_n}$ and observed values $y_n$. Here $\bmath{x_n} = (x_{1,n} ,\ldots, x_{D,n})^T$ is a D-dimensional input vector, and $y_n$ is the corresponding scalar output\footnote{It is straightforward to extend this model to multi-dimensional outputs, but for simplicity we consider only scalars.}. We arrange the $N$ inputs into an $N\times D$ matrix $\mathbfss{X} = (\bmath{x}_1, \ldots, \bmath{x}_N)^T$, and the $N$ observed outputs into a vector $\bmath{y} = (y_1, \ldots, y_N)^T$.

A standard approach is to model data as
\[
\bmath{y} = m(\mathbfss{X},\bphi) + \bepsilon,
\]
where $m$ is the {\it mean function} with parameter vector $\bphi$, and $\bepsilon$ represents independent and identically distributed (i.i.d.) Gaussian noise:
\[
p(\bepsilon) = \mathcal{N}(\bmath{0},\sigma^2\mathbfss{I}).
\]
Here $\mathcal{N}$ represents a multivariate Gaussian distribution with mean vector $\bmath{0}$ and covariance matrix $\sigma^2\mathbfss{I}$, where $\sigma^2$ is the variance of a single observation and and $\mathbfss{I}$ is the $N\times N$ identity matrix. The joint probability distribution for a set of outputs $\bmath y$ is therefore
\[
p(\bmath{y}\,|\,\mathbfss{X},\bphi, \sigma ) = \mathcal{N}(m(\mathbfss{X},\bphi),\sigma^2\mathbfss{I}).
\]
For example to represent photometric observations of a planetary transit, $\bmath{y}$ would be the measured flux, typically the input matrix $\mathbfss{X}$ contains only time, and $\bphi$ are the transit parameters. Thus the above expression is the standard likelihood function often used for transit light curve fitting. This could be extended to include additional inputs in the matrix $\mathbfss{X}$, such as auxiliary data pertaining to the state of the instrument (\eg~detector temperature) and observing conditions (\eg~airmass).

Formally, \citet{Rasmussen_Williams} define a GP as a collection of random variables, any finite number of which have a joint Gaussian distribution. GPs are often written as
\[
y(\bmath{x}) \sim \mathcal{GP} \left (m(\bmath{x}) , \mathbf\Sigma  \right),
\]
where $\mathbf\Sigma$ is the covariance matrix, and we have dropped the mean function parameters from the notation for simplicity. A GP only has two parameters: the mean and covariance. The parameters of the mean function and covariance function,  $\bphi$ and $\btheta$, respectively, are the {\it hyperparameters} of the GP.

As any finite collection of observations has a joint Gaussian distribution, the joint probability distribution of $\bmath{y}$ is given by
\begin{equation}
\label{eq:a1}
p(\bmath{y}\,|\,\mathbfss{X},\bphi, \btheta ) = \mathcal{N}(m(\mathbfss{X}),\mathbf\Sigma).
\end{equation}
We can see that assuming an i.i.d noise model is in fact a special case of a GP with a diagonal covariance matrix. In a GP model, each element in the covariance matrix is defined by the {\it kernel} or {\it covariance function} which has hyperparameters $\btheta$ and is given by
\[
\mathbf\Sigma_{nm} = k(\bmath{x}_n, \bmath{x}_m),
\]
where the kernel maps the input vectors $\bmath{x}$ to a scalar covariance.
The specification of a covariance function implies a probability distribution over functions, thus a GP can be viewed as a distribution over functions. To illustrate this we will adopt a commonly used covariance function, the squared exponential, given by
\[
k(\bmath{x}_n, \bmath{x}_m) = \theta_0 \exp\left(-\theta_1\sum_{i=1}^D(x_{i,n} - x_{i,m})^2\right),
\]
where $\theta_0$ is the maximum covariance and $\theta_1$ is an inverse length scale parameter (this becomes more obvious if we write $\theta_1 = 1/2l^2$ to get an unnormalised Gaussian). As there is only one hyperparameter governing the length scale for all dimensions, this is called an {\it isotropic} kernel.

\begin{figure*}
\includegraphics[width=184mm]{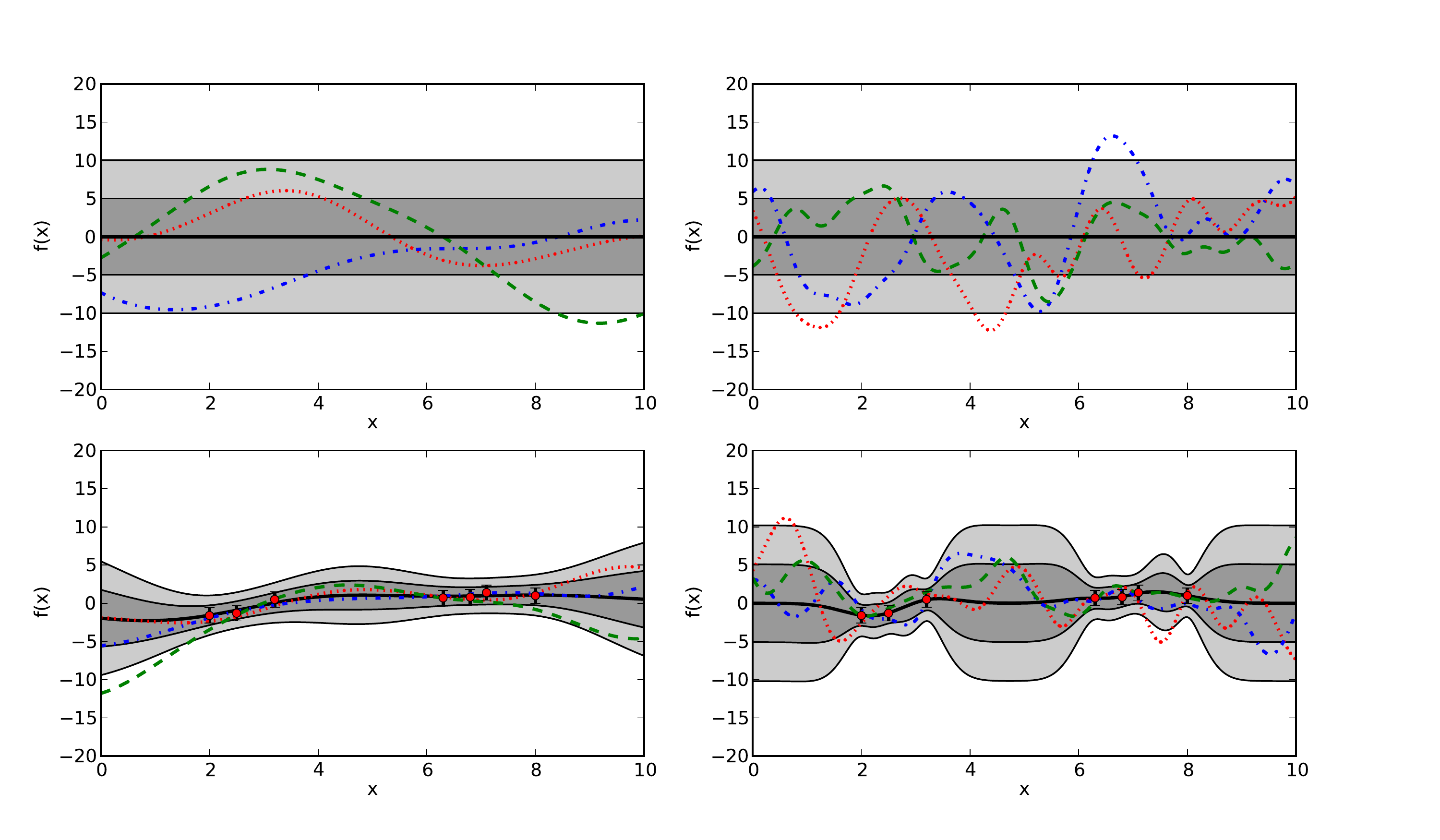}
\caption{Top: Examples of random functions drawn from a GP with a squared exponential covariance kernel. The left and right plots have hyperparameters
$\{\theta_0,\theta_1\} = \{25,2\}$, and $\{\theta_0,\theta_1\} = \{25,0.5\}$, respectively. The dark and light grey shaded areas represent the 1 and 2 $\sigma$ boundaries of the distribution. Bottom: The predictive distributions generated after adding some training data and conditioning using the same hyperparameters as the corresponding plots above, with the addition of a white noise term ($\sigma^2 = 1$). The coloured lines are now random vectors drawn from the posterior distribution.}
\label{fig:gp_regression_1}
\end{figure*}

The GP now has a simple interpretation; data points that lie near each other in input space are highly correlated, and data points which are distant in input space are relatively uncorrelated. We can draw samples from the GP by choosing a number of inputs, calculating the corresponding covariance matrix, and generating a random Gaussian vector from the multivariate distribution. The top two plots in Fig.~\ref{fig:gp_regression_1} show examples of three vectors drawn at random from a GP with a 1 dimensional input space, with a squared exponential kernel function, and zero mean function. The dark and light grey shaded areas represent the 1 and 2$\sigma$ boundaries of the prior distribution, respectively. The left plot has hyperparameters $\{\theta_0,\theta_1\} = \{25,1/8\}$ ($l = 2$), and the right plot has hyperparameters $\{\theta_0,\theta_1\} = \{25,2\}$ ($l = 0.5$). These sample vectors represent random smooth functions drawn from the prior (specified by the hyperparameters). The longer inverse length scale means that the functions on the right plot vary more quickly.

A GP with a squared exponential kernel function therefore defines a distribution over smoothly varying functions $y(\bmath{x})$. The $\theta_0$ hyperparameter determines the maximum variation of the function, and the inverse length scale $\theta_1$ determines how quickly varying the function is; in other words how far input values need to be apart in order  to become uncorrelated. It can be shown that GP regression with the squared exponential kernel function is equivalent to Bayesian linear regression with an infinite number of basis functions\footnote{In practice we need only consider the function values for a finite set of inputs making it possible to work in the infinite space of basis functions. This is impossible using standard basis function models.}.

The mean function of the GP controls the deterministic component of the model, whilst the covariance function controls its stochastic component. Multiple covariance functions can be added or multiplied together to model multiple stochastic processes (as they are still valid covariance functions). For example, most GP models include an i.i.d component. To incorporate this, we simply add a term to the diagonal of the covariance matrix. The covariance function is then given by
\[
k(\bmath{x}_n, \bmath{x}_m) = \theta_0 \exp\left(-\theta_1\sum_{i=1}^D(x_{i,n} - x_{i,m})^2\right) + \delta_{nm} \sigma^2,
\]
where $\delta$ is the Kronecker delta function, and $\sigma^2$ is a new hyperparameter representing the variance of the white noise. Therefore a GP with this covariance function now defines a distribution over smoothly varying functions with the addition of white noise.

The joint probability distribution of the GP given by Eq.~\ref{eq:a1} is the likelihood of observing $\bmath{y}$, given the inputs $\mathbfss{X}$, mean function parameters $\bphi$ and hyperparameters $\btheta$. This likelihood is already marginalised over all possible realisations of the stochastic component that satisfy the covariance matrix (or all possible functions in the distribution specified by the GP) and hence is commonly referred to as the marginal likelihood. In this sense, GPs are intrinsically Bayesian and help to mitigate against the problem of over-fitting associated with non-Bayesian methods.

\subsection{Gaussian Process regression}

Given a set of $N$ observed data points $\bmath{y}$ with corresponding inputs $\mathbfss{X}$ (in machine learning literature known as the training data), we can use a GP to make predictions for additional input values given by $\bmath{x}_\mathbf{\star}$ (known as the test data). We want to obtain the predictive distribution for the output $y_\mathbf{\star}$ conditioned on the training data. According to the prior, the joint probability distribution of the training outputs $\bmath{y}$ and the test output $y_\mathbf{\star}$ is Gaussian and given by
\[
p\left(
\left[
\begin{array}{l} 
\bmath{y}\\y_\mathbf{\star}
\end{array}
\right]
\right)
= \mathcal{N}\left(
\left[
\begin{array}{l} 
m(\mathbfss{X})\\m(\bmath{x}_\mathbf{\star})
\end{array}
\right],
\left[
\begin{array}{cc} 
\mathbf\Sigma & \bmath{k}_\star \\
\bmath{k}_\star^T & c
\end{array}
\right]
\right)
\]
where $\mathbf\Sigma$ is the covariance matrix for the training data points, $\bmath{k}_\mathbf{\star}$ is the column vector with elements $k(\bmath{x}_n, \bmath{x}_\star)$ for $n = (1,\dots,N)$, and $c$ is the scalar given by $k(\bmath{x}_\star, \bmath{x}_\star) + \sigma^2$.

The predictive distribution of $y_\star$ is then obtained by $conditioning$ on the observed data points. Using standard results the conditional joint posterior distribution is another Gaussian:
\[
p\,( y_\star |\, \bmath{x}_\star, \bmath{y}, \mathbfss{X}, \btheta, \bphi ) = \mathcal{N} \left( \hat y_\star, \hat\sigma^2_\star \right),
\]
where $\hat{y}_\star$ and ${\hat\sigma^2}_*$ are the mean and variance of the distribution, respectively, given by
\[
\hat{y}_\star = m(\bmath{x}_\star) + \bmath{k}_\star^T \, \mathbf\Sigma^{-1} \, \bmath{r},
\]
and
\[
{\hat\sigma^2}_* = c - \bmath {k}_\star^T \,\mathbf\Sigma^{-1} \, \bmath {k}_\star,
\]
and $\bmath{r} = \bmath{y} - m(\mathbfss{X})$ is the vector containing the residuals from the mean function.

These results may easily be extended to make predictions for $N_\star$ test inputs given by $\mathbfss{X}_\star$. If we define $\mathbfss{K}_\star$ as the $N \times N_\star$ matrix containing the covariance for all pairs of training inputs and test inputs, and $\mathbfss{K}_{\star\star}$ as the $N_\star\times N_\star$ covariance matrix of the test inputs, the distribution is again Gaussian with mean $\hat{\bmath{y}}$ and covariance matrix $\mathbfss{C}$, given by
\[
\hat{\bmath{y}}_\star = m(\mathbfss{X}_\star) + \mathbfss{K}_\star^T \, \mathbf\Sigma^{-1} \, \bmath{r},
\]
and
\[
\mathbfss{C} = \mathbfss{K}_{\star\star} - \mathbfss{K}_\star^T \,\mathbf\Sigma^{-1} \, \mathbfss{K}_\star.
\]

Fig.~\ref{fig:gp_regression_1} gives simple examples of GP regression conditioned on some artificial training data, shown by the red points. The top plots show the prior probability distribution and random vectors drawn from it, as described in previous section. The lower plots show the distribution conditioned on the training data, with the mean indicated by the thick black line, surrounded by the 1 and 2$\sigma$ predictive distributions. These use the same hyperparameters as the corresponding plots above, with an additional white noise hyperparameter equal to 1. We can think of this process as a weighted average of the functions generated by the prior, weighted on their likelihood function, or mathematically, marginalising over function space. The predictive distribution therefore becomes narrower as more of function space is restricted. Note that with the longer length scale the distribution is narrowed along most of the x range plotted, whereas with the shorter length scale the distribution quickly resorts to the prior distribution far from any observed data. The plotted distributions include white noise for the training data, and the three coloured lines are random functions drawn from the posterior distribution\footnote{When plotting the distribution we need only consider the diagonal of the matrix $\mathbfss{K}_{\star\star}$, whereas when generating random vectors we must consider the full covariance matrix of the test inputs.}. Clearly the value of the hyperparameters has a large impact on the regression; luckily it is possible to infer these from the training data.

\subsection{Learning the hyperparameters}

GPs are extremely effective for interpolation and prediction, and indeed this is one of their main applications in the machine learning community. However, GPs can also be used to infer distributions of the mean function parameters and/or hyperparameters, either of which may be the quantities of interest. We consider the marginal likelihood function from the previous sections:
\[
\mathcal{L}(\mathbf{r}| \mathbfss{X},\btheta,\bphi) = \frac{1}{(2\pi)^{n/2}{| \mathbf{\Sigma}|}^{1/2}}   \exp\left(-\frac{1}{2} \bmath{r}^T\,  \mathbf{\Sigma}^{-1} \, \bmath{r}\right),
\]
where we have now explicitly written the Gaussian function. It is convenient to use the log likelihood:
\[
\log \mathcal{L}(\mathbf{r}| \mathbfss{X},\btheta,\bphi) = -\frac{1}{2} \bmath{r}^T\, \mathbf{\Sigma}^{-1} \, \bmath{r} -\frac{1}{2}\log | \mathbf{\Sigma}| -\frac{n}{2} \log\left(2\pi\right).
\]
Ideally, in a Bayesian framework one should obtain the posterior probability distribution of each hyperparameter of interest by marginalising over all the others. This can be done, for example, using a sampling technique such as Markov Chain Monte Carlo. However, this is expensive to compute as each evaluation of the likelihood function we must invert the $N\times N$ covariance matrix, requiring $\mathcal{O}(n^3)$ operations.

One alternative is to make an approximation known as type-II maximum likelihood. This involves fixing the covariance hyperparameters at their maximum likelihood values, and marginalising over the remaining mean function hyperparameters of interest. This is a valid approximation when the posterior distribution is sharply peaked at its maximum with respect to the covariance hyperparameters, and should give similar results to the full marginalisation.

\begin{figure*}
\includegraphics[width=184mm]{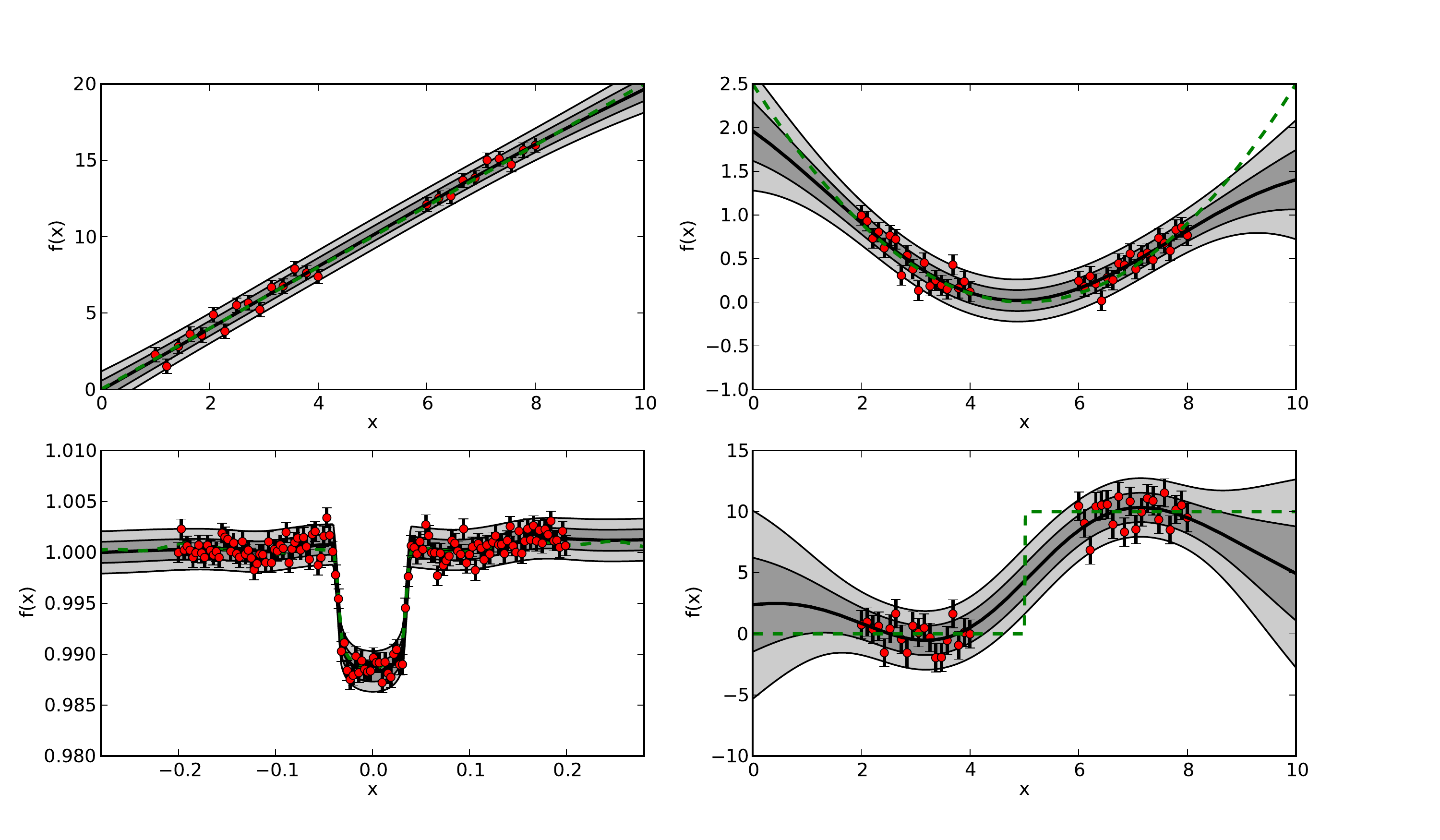}
\caption{Examples of GP regression using type-II maximum likelihood. The dashed green lines represent the functions that the training data were drawn from, before adding Gaussian i.i.d. noise. All data were fitted using a GP with squared exponential kernel, and all used a zero mean function with the exception of the bottom left plot, which used a transit mean function. The solid grey and black regions represent the mean, and 1 and 2$\sigma$ confidence intervals. Clockwise from top left, the function used to generate the data is: a straight line, a quadratic function, a step-function, and a planetary transit model with the addition of correlated 'noise'. See text for discussion.
}
\label{fig:gp_regression_2}
\end{figure*}

Fig.~\ref{fig:gp_regression_2} gives examples of GP regression using a squared exponential covariance function, after optimising the hyperparameters via a Nelder-Mead simplex algorithm \citep[see \eg][]{NumericalRecipes}. The training data (red points) were generated from various functions (shown by the green dashed line) with the addition of i.i.d noise. All were modelled using a zero-mean function, with the exception of the transit function at the bottom left, which used a transit mean function. In the top left the data were generated from a straight line function. The marginal likelihood strongly favours functions with a longer length scale, and the GP is able to predict the underlying function both when interpolating and extrapolating away from the data. The top right data was generated from a quadratic function, and again the GP reliably interpolates the underlying function. However, as the length scale is now shorter, the predictive distribution veers towards the prior more quickly outside the training data. These two examples show that GPs with squared exponential kernels are very good at interpolating data sets providing the data is generated from a smooth function. The bottom left plot shows data drawn from a transit mean function with sinusoidal and Gaussian time-domain `systematics' added. The GP model has a transit mean function, and the additional `systematics' are modelled by the GP. Finally, the bottom right plot shows data drawn from a step function. In this case the GP interpolation is less reliable. This is because the squared exponential function implies smooth functions. A more appropriate choice of covariance kernel is required for such functions \citep[\eg][]{Garnett_2010}. Indeed, a wide choice of covariance kernels exist which effectively model periodic, quasi-periodic and step functions \citep[see][]{Rasmussen_Williams}, enabling GPs to be a very flexible and powerful framework for modelling data in the absence of a parametric model.

\label{lastpage}

\end{document}